# Low-Contact-Resistance Graphene Devices with Nickel-Etched-Graphene Contacts

Wei Sun Leong[†], Hao Gong[‡], and John T. L. Thong[*,†]

[†]Department of Electrical and Computer Engineering, National University of Singapore, Singapore 117583.

[‡]Department of Materials Science & Engineering, National University of Singapore, Singapore 117575.

* Corresponding author's e-mail: elettl@nus.edu.sg

**ABSTRACT**

The performance of graphene-based transistors is often limited by the large electrical resistance across the metal-graphene contact. We report an approach to achieve ultra-low resistance metal contacts to graphene transistors. Through a process of metal-catalysed etching in hydrogen, multiple nano-sized pits with zigzag edges are created in the graphene portions under source/drain metal contacts while graphene channel remains intact. The porous graphene source/drain portions with pure zigzag-termination form strong chemical bonds with the deposited nickel metallization without the need for further annealing. This facile contact treatment prior to electrode metallization results in contact resistance as low as 100 Ωμm in single-layer graphene field-effect transistors, and 11 Ωμm in bilayer graphene transistors. Besides 96% reduction in contact resistance, the contact-treated graphene transistors exhibit 2 folds improvement in mobility. More importantly, the metal-catalysed etching contact treatment is compatible with complementary metal–oxide–semiconductor (CMOS) fabrication processes, and holds great promise to meet the contact performance required for the integration of graphene in future integrated circuits.

Despite the exceptional electronic transport properties of graphene, its potential in electronic devices can only be fully realized if the deleterious effects of parasitic artefacts are addressed. Of these, one of the most serious is the contact resistance which presents a performance bottleneck in graphene transistors.[1-5] To date, studies on electrical contacts to graphene have shown large variations in contact resistance, with reported values as high as hundreds of thousands of $\Omega\mu m$, which are considerably larger than the channel resistance. A common approach to tackle this issue is post-annealing treatment,[6-8] which is known to remove contamination on graphene surfaces,[9] but it is unclear whether the same applies to graphene that had already been covered by metal; indeed Chan *et al.* found that annealing did not significantly affect the contact resistance of their devices.[10] Consequently, a number of alternatives have been explored to minimize contamination at the lithographically defined source/drain contact regions prior to metallization, which include atomic force microscopy (AFM) scanning,[11, 12] ultraviolet/ozone (UVO) treatment[13, 14] and light plasma treatment.[6, 15] However, these approaches are either time-consuming or damaging to the graphene. The question is whether the resulting low contact resistance is due to the removal of resist residues or the creation of defective graphene edges, as the latter establishes "end-contacted" metal-graphene interfaces, which are formed when graphene edges are in contact with metal. Such end-contacts have been predicted to provide much lower contact resistance – up to a few orders of magnitude lower – compared to that of "side-contacted" interfaces.[16] The side-contact configuration is typical of devices fabricated using conventional planar-device fabrication processes, where the metal-graphene interface is dominated by the inert graphene surface, rather than the reactive graphene edges. A good strategy to create end-contacts in a planar-device structure would be to increase the amount of exposed graphene edges at the source/drain regions prior to metallization. As an initial demonstration, Smith *et al.* patterned the source/drain contact regions using electron beam lithography (EBL) and oxygen plasma etching and they observed a 32% of reduction in contact resistance after annealing.[17] However, had the annealing not been applied, they observed an increase in contact resistance instead, as a result of the reduced contact area as well as the structural disorders[18] and amorphization[19] along the plasma-etched graphene edges.

Here, we report an elegant technique that can dramatically increase the amount of defect-free graphene edges exposed at the source/drain contacts. Utilizing a simple Ni-catalysed etching process, a significant amount of etched pits with well-defined zigzag edges is created on the graphene basal plane. The etched pits are formed based on a Ni-catalysed gasification process: C (solid) + 2$H_2$ (gas) → $CH_4$ (gas).[20] We propose the Ni-catalysed etching process as a contact treatment for graphene devices prior to electrode formation. The contact treatment is compatible with CMOS device fabrication processes and holds great potential for the development of CMOS-compatible sub-nanometer graphene devices.

**Results and Discussion**

The contact treatment involves only two steps: metal deposition and annealing. It starts with deposition of Ni at the source/drain regions of graphene device followed by annealing in hydrogen at a moderate temperate for a short duration. Figure 1 illustrates how the contact treatment can be integrated into the fabrication process of a back-gated graphene field-effect transistor (see Methods for details). Ni was chosen as the electrode material for our graphene devices as it is one of the metals that have been predicted to form strong chemical bonds with graphene through orbital hybridization and more importantly, Ni appears to provide the lowest contact resistance to graphene with the smallest variation.[21-24] No further annealing is conducted prior to electrical characterization.

We have been able to observe at least 10 times improvement in contact resistance for our treated single-layer graphene (SLG) device compared against the lowest contact resistance reported previously for Ni-SLG device.[24] For this study, four-point probe measurement technique was used to extract contact resistance of each graphene device *via* equation (1):

$$R_C = \frac{1}{2}(R_{2p} - R_{4p})W \qquad (1)$$

where $R_C$ is the contact resistance, $R_{2p}$ is the device's two-point resistance, $R_{4p}$ is the device's four-point resistance and $W$ is the contact width. Figure 2(a) shows a typical example of a graphene transistor array, which contains a number of graphene transistors with and without Ni-catalysed etching contact treatment. For the untreated devices presented in Figure 2(a), no Ni film was deposited before the annealing step and hence the graphene portion remains intact. The dimensions of all graphene devices made were kept constant. The graphene channel width, channel length, contact width and contact length for all devices are 2 µm. Electrical measurements on all devices were carried out under ambient conditions and all $R_C$ was taken with the back-gate grounded. For this series of devices, the Dirac voltage falls in the range of $V_{Dirac} = (20 \pm 5)$ V.

The measured $R_C$ of more than 40 Ni-contacted graphene devices is plotted in Figure 2(b). For devices with the proposed contact treatment, the average $R_C$ is 89 Ωµm, which is less than 10% of the devices' channel resistance and is about thrice better than the $R_C$ required for state-of-the-art silicon MOSFETs.[25] On the other hand, the average $R_C$ of graphene devices for our untreated Ni-contacted devices is 294 Ωµm, which is about 3 times higher than the average $R_C$ of devices with contact treatment. For comparison purposes, the $R_C$ values reported by others[21, 22] for untreated Ni-contacted exfoliated-graphene devices are also included in Figure 2(b). It is worth noting that the $R_C$ of our graphene devices with contact treatment is about an order of magnitude lower than the $R_C$ values reported by others when Ni is used as electrode metallization.[21, 22] Furthermore, the $R_C$ of our contact-treated graphene devices is not only considerably lower, but also shows a narrower distribution compared to that of untreated devices. The lowest $R_C$ is 11 Ωµm from a contact-treated bilayer graphene device, which represents ~27 times improvement compared to the average $R_C$ of our untreated graphene devices. Remarkably, the smallest $R_C$ of our contact-treated SLG devices is 100 Ωµm which, to the best of our knowledge, is the lowest reported value for SLG devices.

The impact of contact treatment on the mobility extracted from the source/drain terminal characteristics of the field-effect transistor was investigated through back-gate measurements. The parasitic contact resistance degrades the apparent mobility derived from I-V measurements conducted at the device terminals. Figure 3(a) shows the transfer characteristics of three contact-treated graphene field-effect transistors. Each transistor is two-point connected and fabricated *via* the same processes as mentioned above. All transistors were placed in a high vacuum chamber and electrically annealed at 200

°C for 10 hours prior to back-gate measurements. The electrical measurements were carried out at room temperature in vacuum. The peak field-effect mobility was calculated *via* equation (2):

$$\mu_{peak} = \frac{\Delta I_d/V_d \cdot L_{ch}/W_{ch}}{C_{ox} \cdot \Delta V_g} \qquad (2)$$

where *L* and *W* represent channel length and width, respectively, $C_{ox}$ represents the gate capacitance (which is $1.21 \times 10^{-8}$ F/cm$^2$ for 285 nm thick SiO$_2$), $I_d$ is the drain current, $V_d$ is the drain voltage and $V_g$ represents the gate voltage. For the typical contact-treated bilayer graphene transistor in Figure 3(a), the electron mobility is 3916 cm$^2$/V-s, which is 48% better than a previously reported value for an exfoliated bilayer graphene transistor.[26] For fair comparison, we then fabricated two different types of graphene field-effect transistor: contact-treated and untreated. Both graphene devices were made from the same graphene sheet, which has been identified to be a 3-layer graphene and their transfer characteristics are plotted in Figure 3(b). The electron and hole mobilities for the contact-treated graphene transistor are 1818 cm$^2$/V-s and 3579 cm$^2$/V-s, while for the untreated graphene transistor, the electron and hole mobilities are 907 cm$^2$/V-s and 2415 cm$^2$/V-s, respectively. The effective electron and hole mobilities show 2 and 1.5 times improvement, respectively, as a result of reduced contact resistance.

We confirmed the graphene edges formed after the Ni-catalysed etching contact treatment are of pure zigzag configuration through Raman analysis. A bilayer graphene was first patterned into a ribbon using oxygen plasma. Portions of the ribbon were deposited with 2nm of Ni thin film (Figure 4(a)), and the sample underwent annealing in hydrogen following the contact treatment recipe. Raman spectra were obtained at the treated and untreated portions of the graphene ribbon as shown in Figure 4(b). Both treated and untreated portions of graphene have similar Raman spectra and no obvious signal attributable to structural disorders indicating that the proposed treatment does not induce significant defects on the graphene surface. Additionally, Raman spectra at the treated and untreated graphene edges were also acquired at positions indicated in Figure 4(c). According to Figure 4(d), the treated plasma-etched edge has smaller D/G peak intensity ratio and narrower 2D peak (38.96 cm$^{-1}$) when compared to the untreated plasma-etched edge (45.29 cm$^{-1}$). This implies that the treated plasma-etched edge has lower defect density and better atomic crystallinity compared to the untreated plasma-etched edge. Figures 4(e) and (f) show the Raman maps of the intensity of the G-band and the D-band, respectively, of the graphene ribbon. The mapping was performed using a WITecCRM200 Raman system with 532 nm (2.33 eV) excitation with dwell time of 2 seconds and step size of 100 nm. The laser power at the sample was set lower than 0.1 mW to avoid laser-induced heating.[27] The G-band map in Figure 4(e) shows a graphene ribbon structure with uniform intensity. On the other hand, the intensity map of the D-band in Figure 4(f) indicates there are structural defects along the edges of the graphene ribbon. This is not surprising as the graphene ribbon was defined by oxygen plasma initially, which is known to create structural disorders. Remarkably, the intensity of the D-band of treated plasma-etched edges is significantly lower than that of the untreated plasma-etched edges. This is mainly due to the portions of plasma-etched edges having been removed by the Ni-catalysed etching, leaving zigzag edges. Unfortunately, not all disordered structures are etched away as the deposited Ni film is thin, as a result of which it segregates into small particles before the temperature ramps up to the point at which etching initiates. In short, the Raman analysis results corroborate our hypothesis that the proposed contact treatment leaves zigzag edges with low defect density.

We carried out a series of studies to elucidate the morphology after the Ni-catalysed etching contact treatment. As mentioned earlier, the contact treatment involves deposition of thin Ni film on top of graphene surface followed by annealing in a hydrogen environment. The thin Ni film is foreseen to segregate into small particles upon annealing and each particle etches the graphene surface in the presence of hydrogen. The etching process will continue progressively until the Ni front detaches from the graphene edges[20] and finally the Ni balls up leaving behind a triangular or hexagonal etched pit around it. Figure 5(a) shows a typical scanning electron microscope (SEM) image of a few-layer graphene after the treatment where an etch pit could be partially seen under each Ni particle. As SEM imaging of the graphene in the presence of obfuscating Ni particles is not particularly clear, we removed the Ni particles with acid and then characterized the graphene surface using AFM. Figure 5(b) shows a typical AFM image of a treated bilayer graphene. A significant amount of etched pits is observed on the graphene surface and we see evidence of many being triangular in shape, although the surface roughness at this scale makes it difficult to discern clearly. However, the size of different pits varies from 7 nm to 27 nm, with an average of 12 nm. This is due to the tendency of Ni thin film to segregate into islands of different sizes as shown in Figure 5(a). Larger Ni islands etch further before they ball up and such variations give rise to etched pits of different sizes. The inset at the bottom of Figure 5(b) shows the height profile of the marked region while the inset at the top is an enlarged view of a typical triangular pit.

It is worth noting that the size range of etched pits with zigzag edges can be further reduced and it depends on the thickness of the Ni film deposited prior to the annealing process. Thinner films will result in smaller but higher density of etched pits on the graphene surface and *vice versa*. To illustrate this, we repeated the contact treatment process with a 10 nm film of Ni on graphene, which is 5 times thicker than what was presented above. As can be seen, the etched pits are around 500 nm in size, which is much larger and can be easily examined in the SEM (Figure 5(c)). The pits are mostly of hexagonal shape while some are triangular. Figure 5(d) shows two typical large hexagonal etched pits formed on few-layered graphene. One of the etched pits shown still has the Ni adhering to the graphene edges being etched, while another etched pit contains a Ni ball in the middle representing the case of the terminal phase of etching.

We also observed the alignment of etched graphene edges with the Ni lattice in a transmission electron microscope (TEM). Figure 6(a) shows the TEM image of a treated graphene surface comprising different number of graphene layers. Consistent with the SEM observations, the Ni mainly forms into particles on the graphene surface. Along edges of the uppermost graphene layer, we observe Ni particles having etched in from the step (inset of Figure 6(a)) while still being attached to the edge. Nano-beam electron diffraction patterns of a graphene region (position A labeled in Figure 6(a)) and a Ni particle (position B labeled in Figure 6(a)) are shown in Figures 6 (b) and (c), respectively. It should be noted that these patterns are typical, as we observe them from different regions of graphene and different Ni particles. The treated graphene and Ni particle regions have similar crystalline hexagonal symmetry diffraction patterns signifying an epitaxial alignment of Ni (111) with graphene. The diffraction patterns also verify that the treated graphene and Ni particle are both single-crystalline with a lattice spacing of 0.244 nm and 0.246 nm, respectively. Figure 6(d) shows a high resolution TEM image of the dotted square region marked in Figure 6(a). Similar lattice fringes can be seen at both the treated graphene and Ni particle regions. The similarity in lattice fringes is due to less than 1% lattice mismatch between Ni and graphene, which allows a commensurate alignment of Ni with the graphene lattice.[28]

In line with the Ni-catalyzed progressive etching mechanism, the total perimeter of graphene edges exposed at source/drain regions is expected to evolve with the duration of etching. More importantly, the amount of zigzag graphene edges formed at the source/drain contacts has great impact on the amount of end-contacts created in the planar graphene device, which could significantly affect the $R_C$ of graphene devices. Nevertheless, the total perimeter of etched graphene edges will not keep on increasing with the duration of Ni-catalyzed etching, but saturates when the etching discontinues once the Ni detaches and balls up due to surface tension. To investigate the impact of progressive etching mechanism on the $R_C$ of graphene devices, we fabricated a number of devices using the same fabrication processes as illustrated in Figure 1, but the duration of the etching was varied between devices. The contact treatment duration dependence of $R_C$ plotted in Figure S1 agrees well with our hypothesis. The $R_C$ decreases from 570 Ωμm to about 80 Ωμm as the contact treatment duration increases from 2 minutes to 10 minutes and no further reduction is observed beyond 10 minutes of contact treatment.

The contact treatment process creates defect-free zigzag graphene edges that are able to form strong chemical bonds with the subsequent Ni metallization as the metal is deposited. In contrast, the contact area patterning technique presented earlier by Smith et al.[17] results in defective graphene edges at the source/drain regions prior to metallization and 15 hours of high vacuum post-annealing treatment is required to observe the improvement in $R_C$ for graphene devices. To evaluate the effect of post-annealing treatment on our contact-treated graphene devices, we annealed several contact-treated devices in forming gas at 300 ºC for one hour following the first electrical measurement and then repeated the electrical measurement. The extracted $R_C$ values before and after the post-annealing treatment are plotted in Figure S2. It is apparent that the $R_C$ of our contact-treated graphene devices shows minimal improvement after the post-annealing treatment. From a process point of view, this not only saves one process step but also reduces the thermal budget.

**Conclusions**

In summary, we have developed a treatment to improve the metal-graphene contacts through creation of a significant amount of end-contacted graphene edges that are covalently bonded to Ni. Four-point contacted graphene devices with Ni-etched-graphene contacts were fabricated and tested under ambient conditions. The contact-treated graphene devices exhibit $R_C$ as low as 11 Ωµm, with an average of 89 Ωµm, which is ~60% better than the $R_C$ required for silicon MOSFET technology at the 22 nm node.[25] The morphology and chirality of the etched edges have been carefully studied using AFM, SEM, TEM and Raman spectroscopy. The results demonstrate that the proposed Ni-catalyzed etching contact treatment is able to create zigzag graphene edges at the source/drain contact regions and hence allows the formation of strong chemical bonding between metal and graphene. Last but not least, the contact treatment can easily be inserted into a CMOS process flow for future integrated circuits incorporating graphene as an alternative channel material.

**Methods**

**Fabrication of contact-treated graphene field-effect transistors.** To demonstrate the treatment process, graphene flakes were first exfoliated on an oxidized degenerately p-doped silicon substrate with 285 nm thick $SiO_2$. The sample was then spin-coated with a 200 nm thick layer of polymethylmethacrylate (PMMA) 950 A4 (Microchem Inc.) and baked at 120 ºC in an oven for 15 min. Each graphene flake was then delineated into a 2 µm wide ribbon using EBL followed by oxygen plasma etching (20W RF power, 80 V substrate bias, for 30 seconds). Subsequently, the sample was soaked in warm acetone (60 ºC) for more than 12 hours to remove the PMMA layer. After that, a thin film (2 nm) of Ni was deposited at the source/drain contact regions *via* thermal evaporation at a rate of 0.1 nm per second with the channel region protected by a PMMA layer. It was followed by a 12-hour lift-off process in warm acetone. The preceding step was omitted for reference devices (no contact treatment) fabricated on the same graphene flake. Next, the prepared sample was annealed at 580 ºC for half an hour. During annealing, the chamber was filled with a 1:2 mixture of hydrogen and argon at a total gas flow rate of 200 sccm at a pressure of 20 Torr. Finally, the source/drain contacts on graphene were delineated and metallized with 100nm of Ni, without further annealing prior to measurement. For all graphene devices in this work, the dimensions were kept constant. The graphene channel width, channel length, contact width and contact length for all devices are 2 µm as depicted in Figure S4. Ni was chosen as the metallization material because it is one of the metals that have been predicted to form strong chemical bonds with graphene through orbital hybridization. While we chose an annealing temperature of 580 ºC, being the lowest measurable value by an infrared pyrometer, Ni has previously been shown to etch graphite surfaces at 550 ºC.[29] Higher temperatures (>1000 ºC) should be avoided to prevent nanoparticle etching[30] from taking place, which would otherwise cut swathes across the graphene.

**References and Notes**

1. Novoselov, K. S.; Falko, V. I.; Colombo, L.; Gellert, P. R.; Schwab, M. G.; Kim, K. A Roadmap for Graphene. *Nature* 2012, 490, 192-200.
2. Liu, W.; Wei, J.; Sun, X.; Yu, H. A Study on Graphene—Metal Contact. *Crystals* 2013, 3, 257-274.
3. Schwierz, F. Graphene Transistors. *Nat. Nanotechnol.* 2010, 5, 487-496.
4. Xia, F.; Perebeinos, V.; Lin, Y. M.; Wu, Y.; Avouris, P. The Origins and Limits of Metal-Graphene Junction Resistance. *Nat. Nanotechnol.* 2011, 6, 179-184.
5. Wang, L.; Meric, I.; Huang, P. Y.; Gao, Q.; Gao, Y.; Tran, H.; Taniguchi, T.; Watanabe, K.; Campos, L. M.; Muller, D. A. *et al.* One-Dimensional Electrical Contact to a Two-Dimensional Material. *Science* 2013, 342, 614-617.
6. Robinson, J. A.; LaBella, M.; Zhu, M.; Hollander, M.; Kasarda, R.; Hughes, Z.; Trumbull, K.; Cavalero, R.; Snyder, D. Contacting Graphene. *Appl. Phys. Lett.* 2011, 98, 053103-1-3.
7. Balci, O.; Kocabas, C. Rapid Thermal Annealing of Graphene-Metal Contact. *Appl. Phys. Lett.* 2012, 101.
8. Malec, C. E.; Elkus, B.; Davidović, D. Vacuum-Annealed Cu Contacts for Graphene Electronics. *Solid State Commun.* 2011, 151, 1791-1793.
9. Lin, Y.-C.; Lu, C.-C.; Yeh, C.-H.; Jin, C.; Suenaga, K.; Chiu, P.-W. Graphene Annealing: How Clean Can It Be? *Nano Lett.* 2011, 12, 414-419.
10. Chan, J.; Venugopal, A.; Pirkle, A.; McDonnell, S.; Hinojos, D.; Magnuson, C. W.; Ruoff, R. S.; Colombo, L.; Wallace, R. M.; Vogel, E. M. Reducing Extrinsic Performance-Limiting Factors in Graphene Grown by Chemical Vapor Deposition. *ACS Nano* 2012, 6, 3224-3229.
11. Goossens, A. M.; Calado, V. E.; Barreiro, A.; Watanabe, K.; Taniguchi, T.; Vandersypen, L. M. K. Mechanical Cleaning of Graphene. *Appl. Phys. Lett.* 2012, 100, 073110-1-3.
12. Lindvall, N.; Kalabukhov, A.; Yurgens, A. Cleaning Graphene Using Atomic Force Microscope. *J. Appl. Phys.* 2012, 111, 064904-1-4.
13. Chen, C. W.; Ren, F.; Chi, G. C.; Hung, S. C.; Huang, Y. P.; Kim, J.; Kravchenko, II; Pearton, S. J. UV Ozone Treatment for Improving Contact Resistance on Graphene. *J. Vac. Sci. Technol. B* 2012, 30, 060604-1–3.
14. Li, W.; Liang, Y.; Yu, D.; Peng, L.; Pernstich, K. P.; Shen, T.; Walker, A. R. H.; Cheng, G.; Hacker, C. A.; Richter, C. A. *et al.* Ultraviolet/Ozone Treatment to Reduce Metal-Graphene Contact Resistance. *Appl. Phys. Lett.* 2013, 102, 183110-1-5.
15. Choi, M. S.; Lee, S. H.; Yoo, W. J. Plasma Treatments to Improve Metal Contacts in Graphene Field Effect Transistor. *J. Appl. Phys.* 2011, 110, 073305-1–6.
16. Matsuda, Y.; Deng, W.-Q.; Goddard, W. A. Contact Resistance for "End-Contacted" Metal−Graphene and Metal−Nanotube Interfaces from Quantum Mechanics. *J. Phys. Chem. C* 2010, 114, 17845-17850.
17. Smith, J. T.; Franklin, A. D.; Farmer, D. B.; Dimitrakopoulos, C. D. Reducing Contact Resistance in Graphene Devices through Contact Area Patterning. *ACS Nano* 2013, 7, 3661-3667.
18. Han, M. Y.; Özyilmaz, B.; Zhang, Y.; Kim, P. Energy Band-Gap Engineering of Graphene Nanoribbons. *Phys. Rev. Lett.* 2007, 98, 206805.
19. Ryu, S.; Maultzsch, J.; Han, M. Y.; Kim, P.; Brus, L. E. Raman Spectroscopy of Lithographically Patterned Graphene Nanoribbons. *ACS Nano* 2011, 5, 4123-4130.
20. Wang, R.; Wang, J.; Gong, H.; Luo, Z.; Zhan, D.; Shen, Z.; Thong, J. T. L. Cobalt-Mediated Crystallographic Etching of Graphite From Defects. *Small* 2012, 8, 2515-2523.
21. Nagashio, K.; Nishimura, T.; Kita, K.; Toriumi, A. Metal/Graphene Contact as a Performance Killer of Ultra-High Mobility Graphene - Analysis of Intrinsic Mobility and Contact Resistance. *IEDM Tech. Dig.* 2009, 565–568.
22. Venugopal, A.; Colombo, L.; Vogel, E. M. Contact Resistance in Few and Multilayer Graphene Devices. *Appl. Phys. Lett.* 2010, 96, 013512-1-3.

23. Russo, S.; Craciun, M. F.; Yamamoto, M.; Morpurgo, A. F.; Tarucha, S. Contact Resistance in Graphene-Based Devices. *Phys. E (Amsterdam, Neth.)* 2010, 42, 677-679.
24. Nagashio, K.; Nishimura, T.; Kita, K.; Toriumi, A. Contact Resistivity and Current Flow Path at Metal/Graphene Contact. *Appl. Phys. Lett.* 2010, 97, 143514-1-3.
25. International Technology Roadmap for Semiconductors. In *Process Integration, Devices, and Structures* 2012. http://www.itrs.net/Links/2012ITRS/Home2012.htm.
26. Franklin, A. D.; Han, S. J.; Bol, A. A.; Perebeinos, V. Double Contacts for Improved Performance of Graphene Transistors. *IEEE Electron Device Lett.* 2012, 33, 17-19.
27. Hao, Y.; Wang, Y.; Wang, L.; Ni, Z.; Wang, Z.; Wang, R.; Koo, C. K.; Shen, Z.; Thong, J. T. L. Probing Layer Number and Stacking Order of Few-Layer Graphene by Raman Spectroscopy. *Small* 2010, 6, 195-200.
28. Mattevi, C.; Kim, H.; Chhowalla, M. A Review of Chemical Vapour Deposition of Graphene on Copper. *J. Mater. Chem.* 2011, 21, 3324-3334.
29. Keep, C. W.; Terry, S.; Wells, M. Studies of The Nickel-Catalyzed Hydrogenation of Graphite. *J. Catal.* 1980, 66, 451-462.
30. Campos, L. C.; Manfrinato, V. R.; Sanchez-Yamagishi, J. D.; Kong, J.; Jarillo-Herrero, P. Anisotropic Etching and Nanoribbon Formation in Single-Layer Graphene. *Nano Lett.* 2009, 9, 2600-2604.

FIGURES

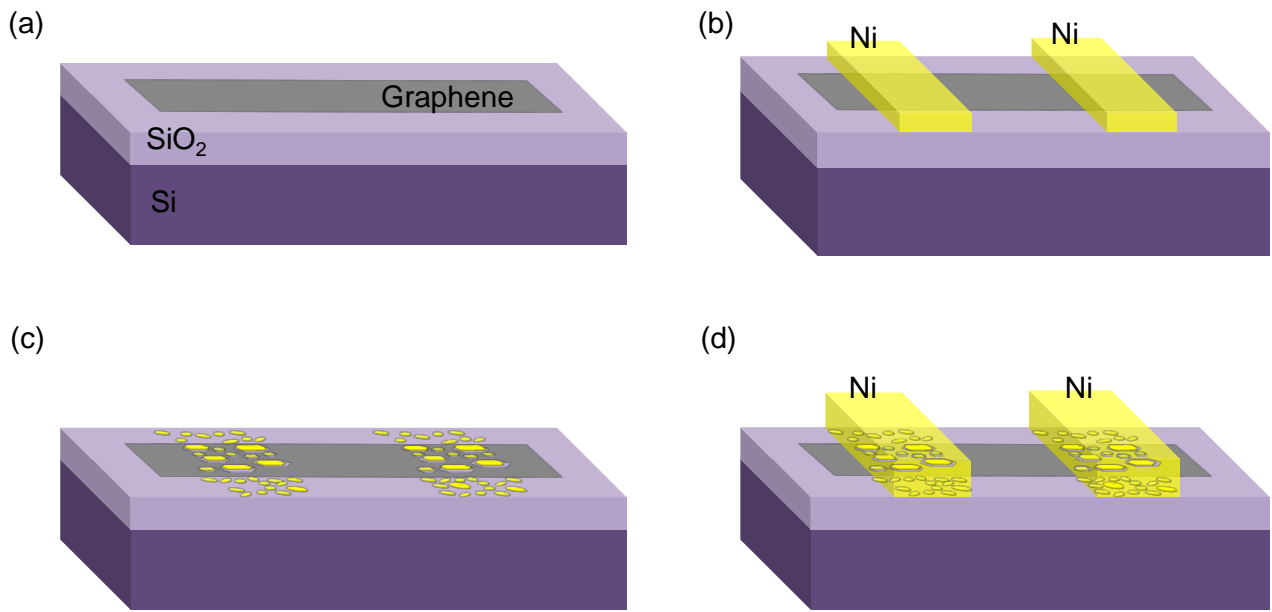

**Figure 1.** Schematics of the process showing the fabrication steps of a back-gated graphene field-effect transistor with Ni-etched-graphene contacts. (a) Exfoliated graphene on a $p^+$ Si / $SiO_2$ substrate is patterned into a strip using electron beam lithography and oxygen plasma etching. (b) Thin Ni films are deposited at the source/drain regions. (c) After annealing in hydrogen, large amount of pits enclosed by zigzag graphene edges formed within the source/drain regions. (d) Thick Ni metallization deposited as electrical contacts to the graphene device forming Ni-etched-graphene contacts.

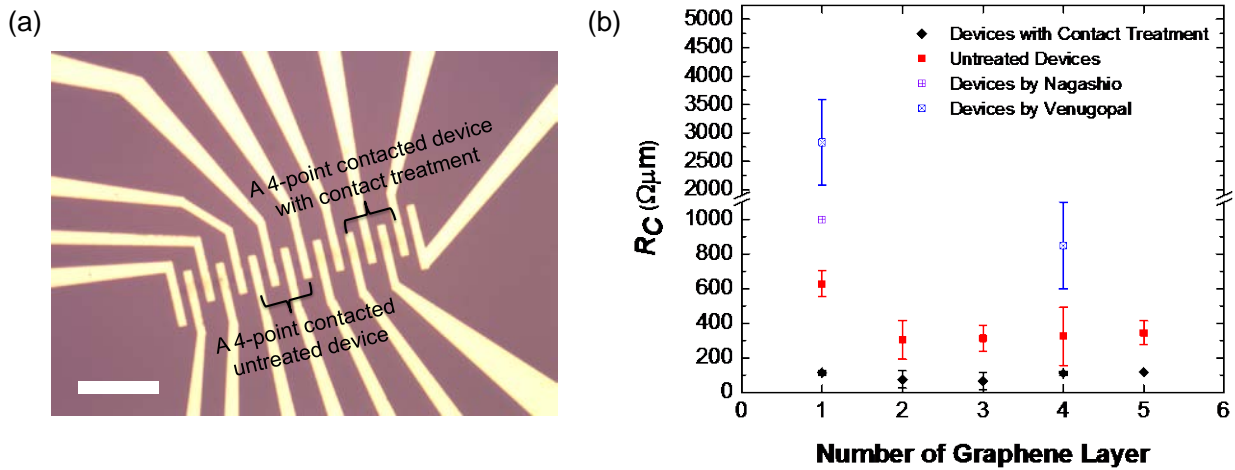

**Figure 2.** Contact resistance comparison for graphene devices with and without contact treatment. (a) An array of graphene transistors fabricated with the proposed process flow. Scale bar: 20 µm. (b) Contact resistance distribution of Ni-contacted graphene devices. The average $R_C$ of our contact-treated graphene devices is 89 Ωµm, which is ~3 folds better than that of untreated devices (294 Ωµm), with the lowest of 11 Ωµm from a bilayer graphene device (>27 times of improvement compared to the average $R_C$ of our untreated graphene devices). The $R_C$ values reported by Nagashio[21] and Venugopal[22] for Ni-contacted exfoliated-graphene devices are included for comparison.

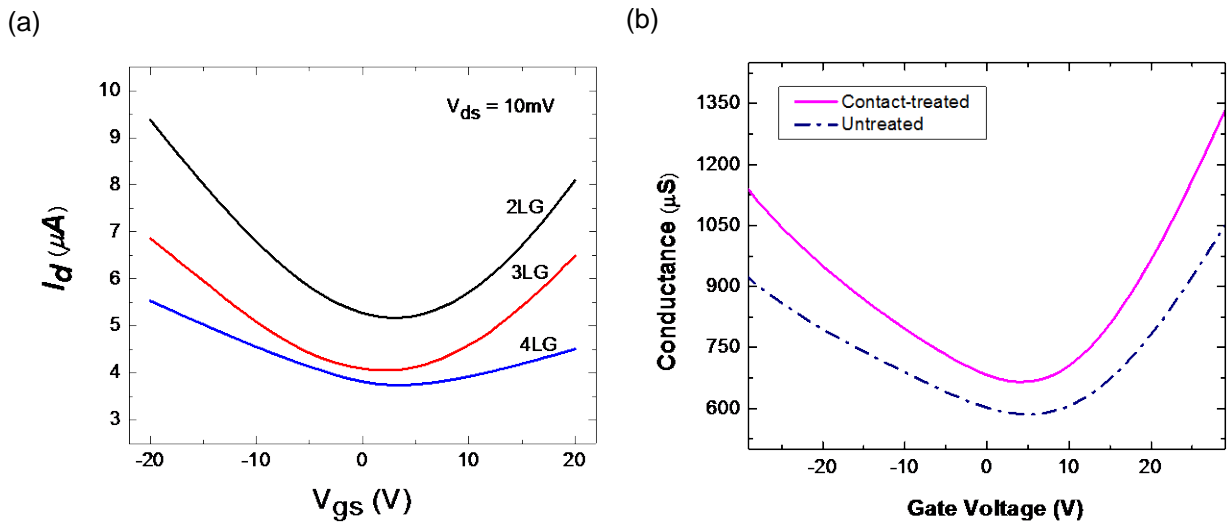

**Figure 3.** Back-gate measurement results on graphene devices with and without contact treatment. (a) Transfer characteristics of 3 contact-treated graphene field-effect transistors with different number of graphene layers. The field-effect electron mobility for the contact-treated bilayer graphene transistor is 3916 cm$^2$/V-s, which is 48% better than the previously reported value.[24] (b) Both contact-treated and untreated graphene field-effect transistors were fabricated from the same 3-layer graphene sheet for fair comparison. The effective electron and hole mobilities show 2 and 1.5 times improvement, respectively, as a result of reduced contact resistance.

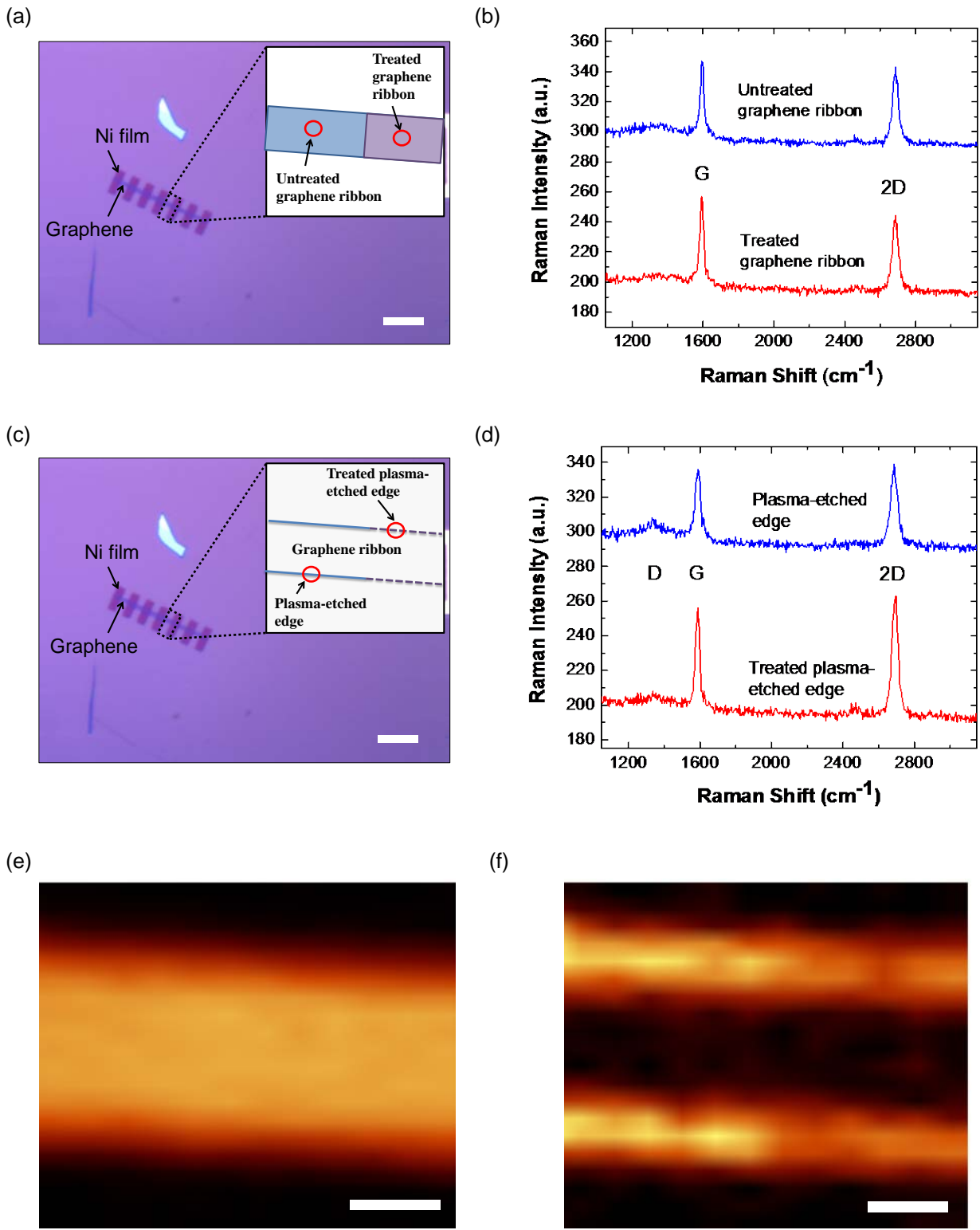

**Figure 4.** Verification of zigzag graphene edges through Raman analysis. (a) Optical image of a graphene ribbon defined by oxygen plasma and then partially treated with Ni-catalyzed etching. Scale bar: 10 µm. Inset: Schematic of the indicated position. (b) Raman spectrum taken at the positions

indicated in 5(a). Both treated and untreated portions of graphene have similar Raman spectra and contain no D-band signal. (c) Optical image of a graphene ribbon defined by oxygen plasma and then partially treated with etching. Scale bar: 10 μm. Inset: Schematic of a graphene ribbon with the difference along edges labeled. (d) Raman spectrum taken at the positions indicated in (c). The treated plasma-etched edge has smaller D/G peak intensity ratio and narrower 2D peak compared to the untreated plasma-etched edge. (e) Raman maps showing the intensity of G-band, and (f) D-band of the particular portion of the graphene ribbon as indicated by the dotted rectangle in (c). Scale bars: 500 nm. The graphene ribbon shows no intensity difference in G-band signal, but obvious difference in D-band signal between the treated and untreated plasma-etched edges.

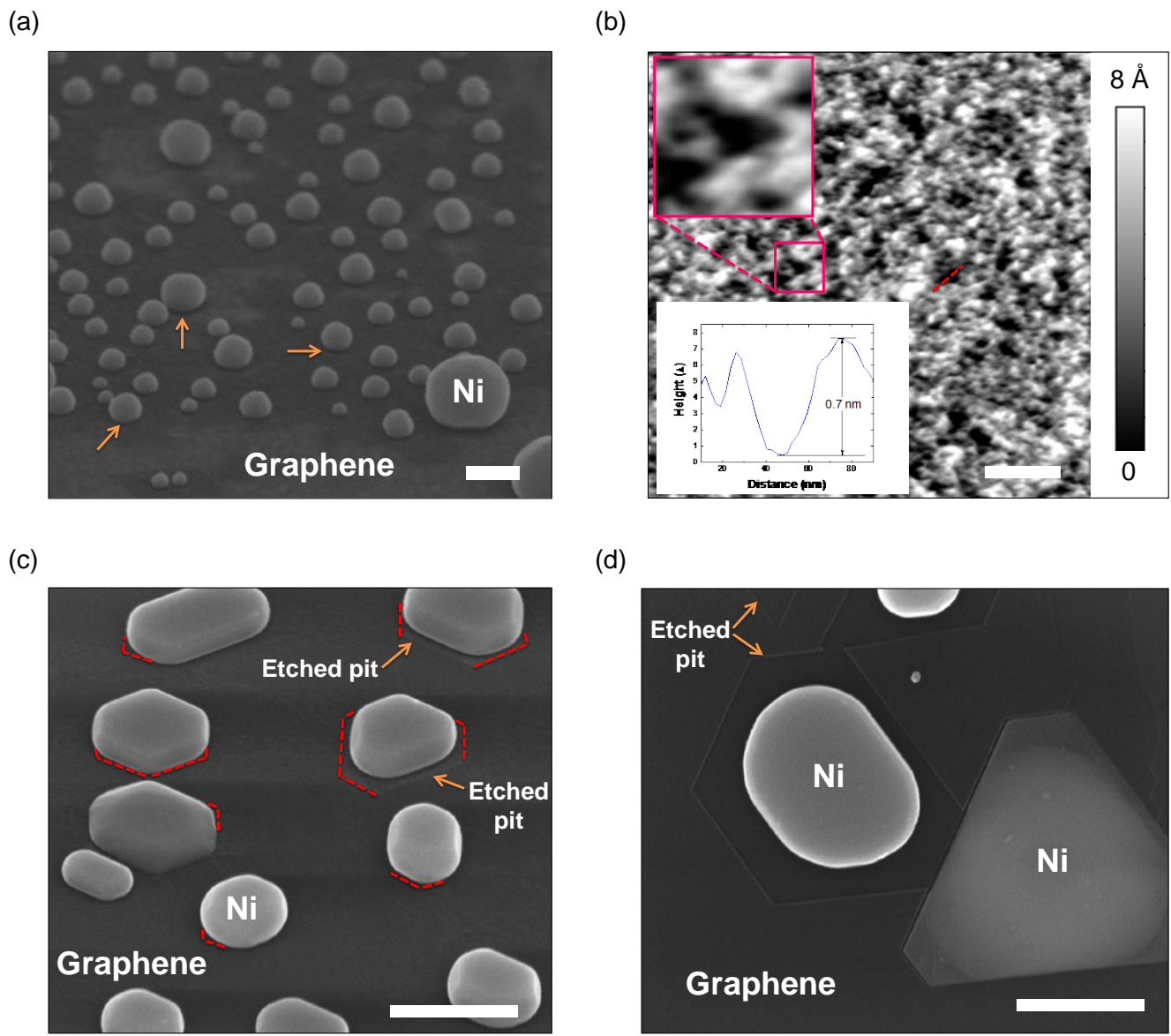

**Figure 5.** SEM and AFM characterization of graphene surface after contact treatment. (a) Typical 45° tilted SEM image of etched few-layer graphene showing balled-up Ni sitting in the middle of each etched pit. Arrows indicates some etched pits that can be partially seen. Scale bar: 100 nm. (b) Typical AFM image of a bilayer graphene after removal of Ni balls. Scale bar: 100 nm. Insets: A typical triangular etched pit (top) and height profile along the dotted line (bottom). (c) Typical 45° tilted SEM image of etched few-layer graphene surface using thicker Ni film, resulting in larger and more visible etched pits. Scale bar: 500 nm. (d) Typical SEM image showing top view of two large hexagonal etched pits on few-layer graphene. One of etched pits contains a Ni ball in the middle representing the case of the terminal phase of etching, while another etched pit shown still has the Ni adhering to the graphene edges being etched. Scale bar: 500 nm.

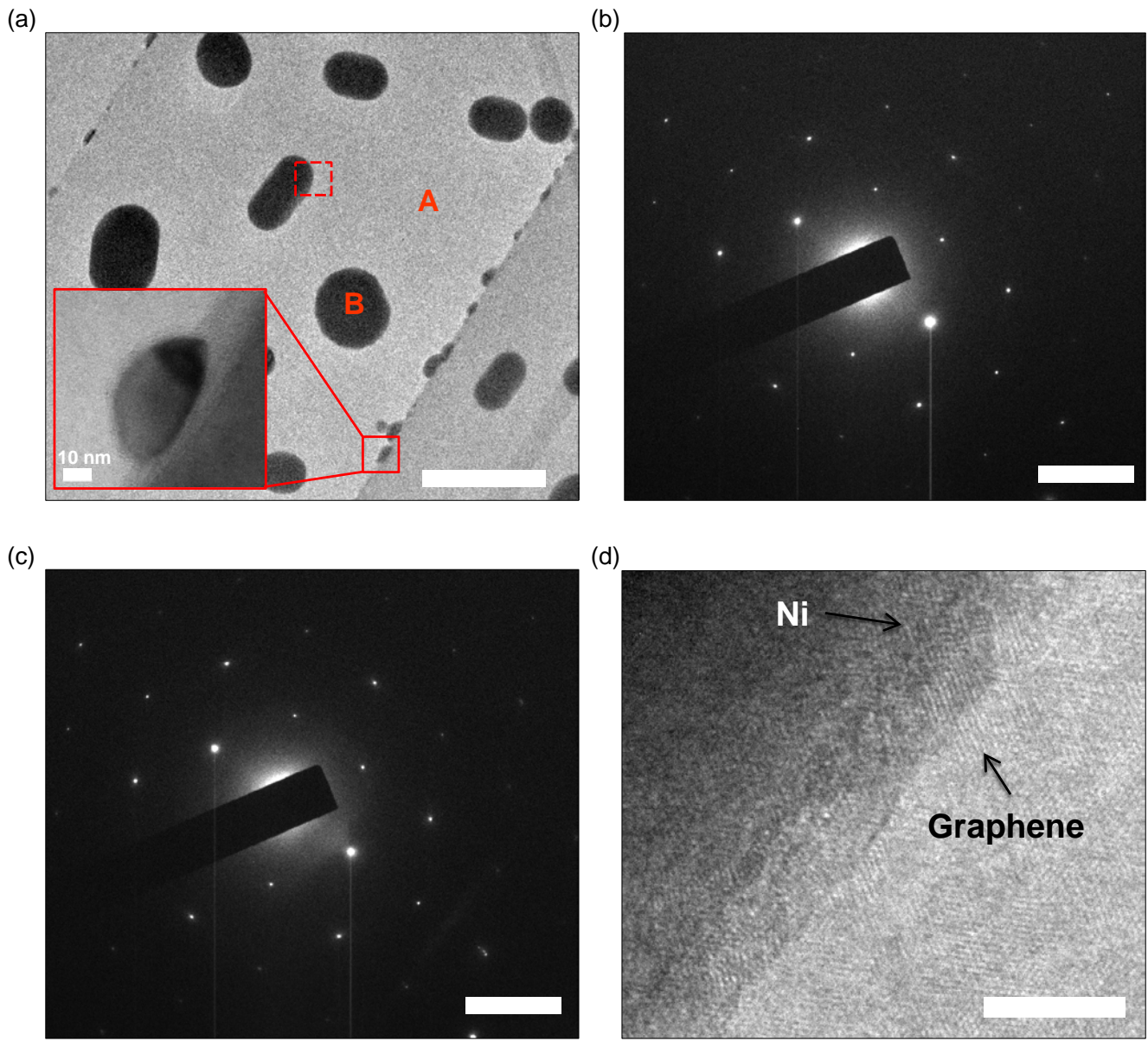

**Figure 6.** TEM characterization of graphene surface after contact treatment. (a) A TEM image of treated graphene surface. Scale bar: 500 nm. The inset shows a Ni particle residing in the graphene edges. (b) Hexagonal electron diffraction pattern of graphene region (position A indicated in (a)). Scale bar: 51 nm$^{-1}$. (c) Hexagonal electron diffraction pattern of a Ni particle (position B indicated in (a)). Scale bar: 51 nm$^{-1}$. (d) The high resolution TEM image of the region contained within the dotted square indicated in (a). Similar lattice fringes observed at the Ni-graphene interface. Scale bar: 5 nm.

# APPENDIX

## S1. Investigation of the impact of progressive etching mechanism on the $R_C$ of graphene devices

A number of graphene transistors were made using the same fabrication processes as illustrated in the main text, but the duration of the etching process was varied between transistors. Figure S1 shows the treatment duration dependence of $R_C$, which decreases from 570 Ωμm to about 80 Ωμm as the duration increases from 2 minutes to 10 minutes. No further reduction in $R_C$ was observed beyond 10 minutes of contact treatment. This observation is in line with a mechanism of progressive etching that increases the total perimeter of graphene edges, until the point where the metal front detaches from the graphene edge due to surface tension effects that finally cause the Ni to ball up.

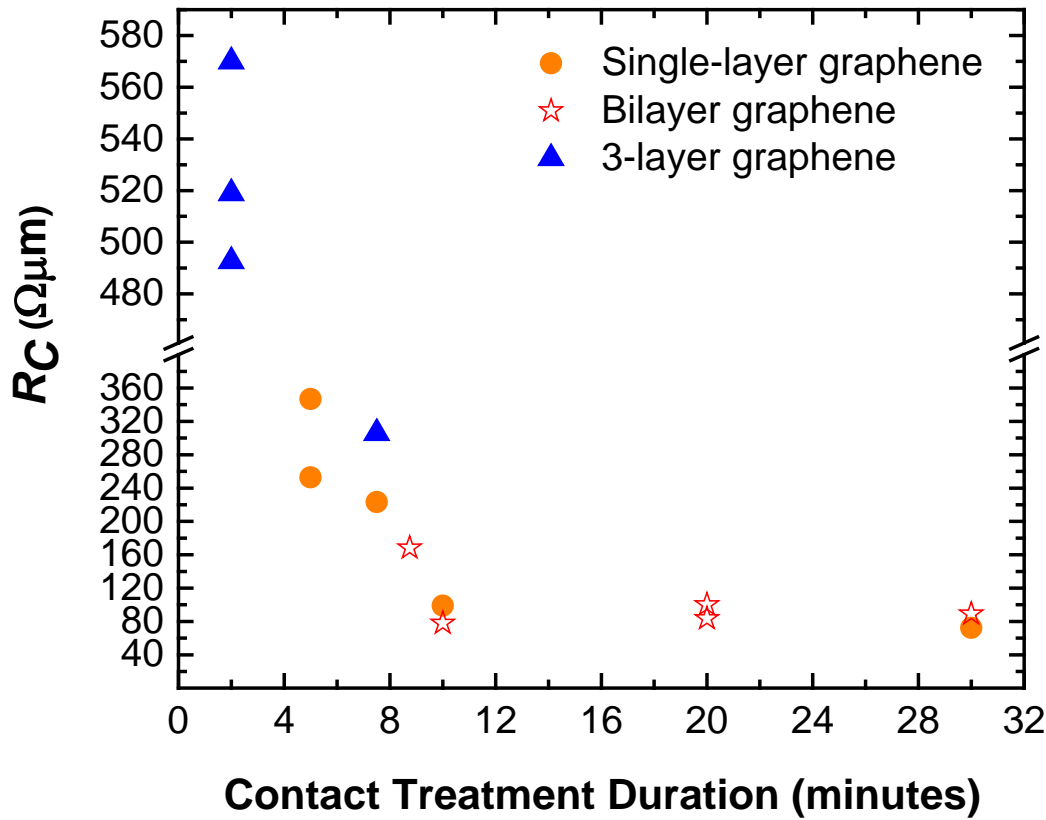

**Figure S1.** Measured contact resistance of contact-treated graphene transistors as a function of the Ni-mediated etching contact treatment duration.

## S2. Effect of post-annealing treatment on the Ni-catalysed etching treated graphene devices

We also evaluated the effect of post-annealing treatment on the contact-treated graphene devices. Following the first electrical measurement, several devices with contact treatment were annealed in forming gas at 300°C for one hour. The electrical measurement was repeated and the extracted $R_C$ values are plotted in Figure S2. The $R_C$ of each device does reduce but the improvement is not significant. This is not unexpected as the graphene edges created through the Ni-mediated etching are mostly zigzag edges, which form strong chemical bonds with the subsequent Ni metallization as the metal is deposited.

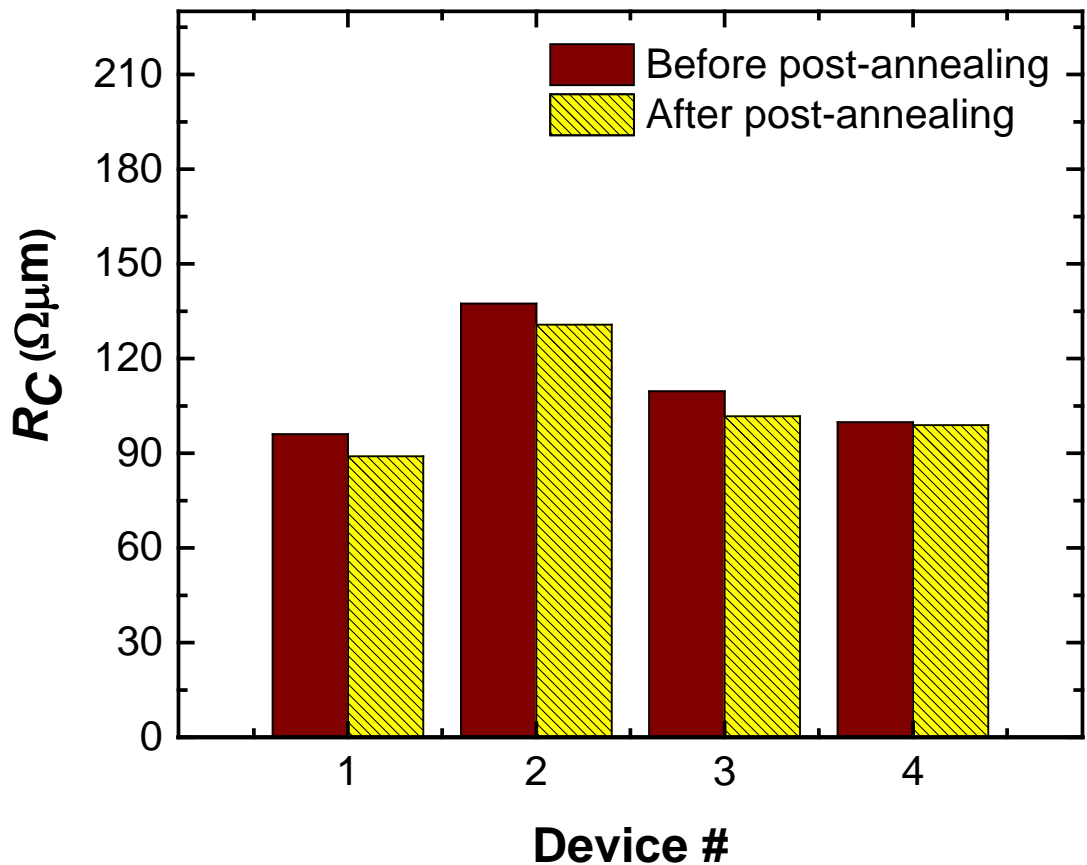

**Figure S2.** Measured contact resistance of contact-treated graphene transistors before and after post-annealing treatment.

## S3. Mechanism of pit formation in graphene

The contact treatment process was repeated with a 10 nm film of Ni on graphene and 580 °C annealing in high vacuum ($10^{-6}$ mbar) to confirm the proposed Ni-catalysed etching mechanism. As can be seen in Figure S3, the Ni particles formed on few-layer graphene surface are around 500 nm in size, which is similar to the situation observed in Figure 5(c). However, no pit is formed on the graphene surface in this case (hydrogen is absent during the annealing process). This result confirms that the pits formed in the graphene are based on the catalytic gasification process of carbon in a hydrogen atmosphere forming methane as follows: C (solid) + $2H_2$ (gas) → $CH_4$ (gas).

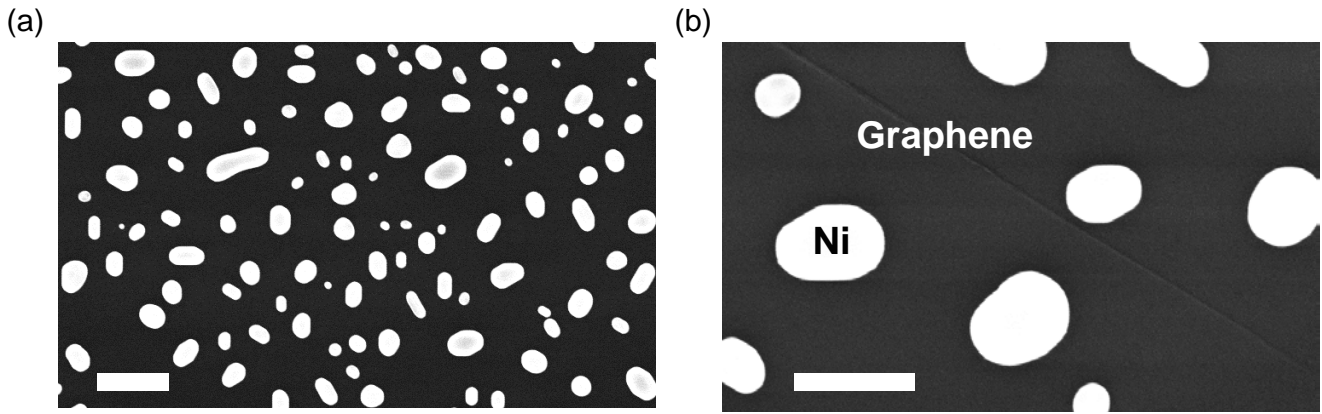

**Figure S3.** SEM characterization of graphene surface after contact treatment in vacuum. (a) Typical 45° tilted SEM image of a few-layer graphene surface using 10 nm of Ni film. Scale bar: 1 µm. (d) Typical 45° tilted higher magnification SEM image showing Ni particles on few-layer graphene without any etching phenomenon observed. Scale bar: 500 nm.